\documentclass[conference]{IEEEtran}
\IEEEoverridecommandlockouts
\usepackage{cite}
\usepackage{amsmath,amssymb,amsfonts}
\usepackage{algorithmic}
\usepackage{graphicx}
\usepackage{textcomp}
\usepackage{xcolor}
\usepackage{soul}        
\usepackage{array}
\usepackage{multirow}
\usepackage{tabularx}
\usepackage{adjustbox}
\usepackage{framed,enumitem}
\usepackage{amssymb}
\usepackage{hyperref}
\usepackage{array}
\usepackage{booktabs}
\usepackage{csquotes}
\usepackage[utf8]{inputenc}
\newlist{todolist}{itemize}{2}
\setlist[todolist]{label=$\square$}
\usepackage{pifont}
\usepackage[T1]{fontenc}

\def\BibTeX{{\rm B\kern-.05em{\sc i\kern-.025em b}\kern-.08em
    T\kern-.1667em\lower.7ex\hbox{E}\kern-.125emX}}

\begin{document}

\title{Non-functional Requirements for Machine Learning:  Understanding Current Use and Challenges in Industry\\
}

\author{\IEEEauthorblockN{Khan Mohammad Habibullah, Jennifer Horkoff}
\IEEEauthorblockA{\textit{Chalmers $|$ University of Gothenburg, Sweden} \\
\{khan.mohammad.habibullah, jennifer.horkoff\}@gu.se}

}

\maketitle

\begin{abstract}
Machine Learning (ML) is an application of Artificial Intelligence (AI) that uses big data to produce complex predictions and decision-making systems, which would be challenging to obtain otherwise. To ensure the success of ML-enabled systems, it is essential to be aware of certain qualities of ML solutions (performance, transparency, fairness), known from a Requirement Engineering (RE) perspective as non-functional requirements (NFRs). 
However, when systems involve ML, NFRs for traditional software may not apply in the same ways; some NFRs may become more prominent or less important; NFRs may be defined over the ML model, data, or the entire system; and  NFRs for ML may be measured differently.  In this work, we aim to understand the state-of-the-art and challenges of dealing with NFRs for ML in industry.  We interviewed ten engineering practitioners working with NFRs and ML. We find examples of (1) the identification and measurement of NFRs for ML, (2) identification of more and less important NFRs for ML, and (3) the challenges associated with NFRs and ML in the industry.  This knowledge paints a picture of how ML-related NFRs are treated in practice and helps to guide future RE for ML efforts.

\end{abstract}


\vspace{-0.2cm}
\section{Introduction} \label{sec:intro}


Machine Learning (ML) is  increasingly used in decision-making and prediction applications, including image recognition,  language processing, and autonomous systems. ML includes algorithms that use large amounts of data to learn and automatically perform tasks that are challenging with traditional software~\cite{smola2008introduction}. ML-enabled software often has an influence on critical decision making (e.g., cancer detection and loan approval), and such decisions may suffer from unintended bias~\cite{kamishima2011fairness}, unsafe execution~\cite{challen2019artificial}, unexplainable decisions~\cite{biran2017explanation}. At the same time, ML-enabled software is becoming more complex, and exhaustive testing is expensive and time-consuming. Because of these issues, ensuring safe, fair, transparent, and accurate ML-enabled  systems is challenging. 
From a Requirement Engineering (RE) perspective, these quality aspects are  known as non-functional requirements (NFRs)~\cite{chung2012non}. 
 
 For more than 40 years, work has focused on how to consider and deal with software qualities or NFRs in an effective way as part of RE and software development, e.g.,~\cite{cavano1978framework}. Although much work has been devoted to NFRs, including exploring effective definitions of this concept~\cite{glinz2007non}, dealing with NFRs remains a difficult challenge in modern system development~\cite{NFRAgileSurvey}. Glinz defines NFRs as ``an attribute of or a constraint on a system'', where attributes are performance or quality requirements~\cite{glinz2007non}. In this work, we focus on NFRs as quality requirements.  Despite the NFR-related challenges, progress in the area of NFR exploration has been made, including, for example, definitions (e.g.,~\cite{glinz2007non}), taxonomies (e.g.,~\cite{galster2008taxonomy}), classification methods (e.g.,~\cite{cleland2007automated}), modeling approaches (e.g.,~\cite{chung2012non}), management methods (e.g.,~\cite{kaur2014non}), and industrial studies (e.g.,~\cite{eckhardt2016non}).  
 
 However, when dealing with ML-enabled software, it is not clear if our accumulated knowledge concerning NFRs is still applicable.
  Looking at the literature (e.g.,~\cite{kamishima2011fairness}), popular science (e.g.,~\cite{Obermeyer447}), and reading AI-related news~\cite{reutersprivacy}, we can see that some NFRs, such as fairness, become more prominent, others, such as privacy, remain relevant, and  perhaps others, such as usability, become less relevant.  
  Further, as-yet-unexplored NFRs such as “retrainability” may also become relevant.  But does knowledge gained from available literature reflect current practice? In order to begin to answer how our knowledge of NFRs should be reconsidered, we need to understand the state of practice concerning NFRs and ML in industry.

Existing work has begun to look at  challenges with ML use in practice. According to a recent survey, RE is the most challenging activity for ML-enabled software development~\cite{ishikawa2019engineers}. 
Several recent papers have identified and discussed RE-related challenges in AI-based systems ~\cite{belani2019requirements,Heyn2021,vogelsang2019requirements}. 
However, we are unable to find work which  focuses explicitly on how NFRs are dealt with in an industrial ML-context. 


 NFRs can be defined at over different granularity levels of a system, i.e., we can define NFRs for a system, a component, or a specific feature.  Machine Learning is often only a small part of a larger system~\cite{sculley2015hidden}. 
 NFRs can be identified and measured over ML-related data, over the  ML-model, or over the whole software which includes ML. In which of these areas do we see changes in how NFRs are identified, managed or measured?   Are these distinctions in scope used in practice? Where do the NFR-related challenges lie?  
To begin to address these and other questions,
we conducted interviews with ten practitioners who have worked with ML in  industry,  exploring their perceptions of NFRs in an ML context. The interview covered more or less important NFRs for ML-enabled software, how NFRs are captured and measured, and what challenges they face to working with and measuring NFRs for ML. 

Our interviewees agree that NFRs for ML are important and play a significant role in the software's overall success. They could identify many important NFRs for ML-enabled systems and some tools and techniques to measure NFRs in the ML industry. The industry practitioners also brought up challenges encountered in  identifying and measuring NFRs for ML-enabled software.  However, we find that currently, industry cannot offer many extensive or well-established solutions in how to identify, scope, or measure NFRs for ML. 


The contributions of this paper including raising awareness of NFRs in ML amongst practitioners, 
bringing NFRs for ML to the forefront, and facilitating early consideration. 
From a research perspective, we identify a number of challenges which can direct future work in the area of NFRs for ML systems, including studying particular NFRs, considering the scope of NFRs, and identifying ML-specific NFR measurements.

The  paper is structured as follows: Sec. \ref{Related Work} describes related work, while Sec. \ref{Research Questions} states the research questions. We describe the research method in Sec. \ref{Methodology}, and present our findings related to NFRs in Sec. \ref{Results}. Sec. \ref{Discussion} discusses our findings and future work. We conclude our study in Sec. \ref{Conclusions}.
\vspace{-0.1cm}
\section{Related Work} \label{Related Work}

In this section, we highlight relevant related work topics, including NFRs, RE for AI, and work on SE for AI.

\textbf{NFRs.}  Although NFRs are considered essential and critical for  software success, there is no agreed upon guidelines on when and how NFRs should be elicited, documented, and validated~\cite{glinz2007non}. 
Even so, as highlighted in Sec.~\ref{sec:intro}, much work in an out of RE has been devoted to NFRs, so much so that it is difficult to give a brief and complete overview.   

While many papers looked into the NFRs for traditional software, a few studies  focused on NFRs for ML-enabled systems in industry. Doerr et al. presented the application of a systematic, experience-based method to elicit, document, and analyze non-functional requirements. Their objective was to achieve a sufficient set of measurable and traceable non-functional requirements~\cite{doerr2005non}. Ameller et al. conducted an interview-based survey with 18 different companies from six European countries. They presented the barriers to and benefits of the management of NFRs for companies, how NFRs are supported by Model-Driven Development (MDD), and which strategies are followed if some NFRs are not supported by MDD approaches.  Their results show that MDD adaptation is a complex process with little or no support for NFRs, and productivity and maintainability should be supported when MDD is adopted~\cite{ameller2019dealing}. Sachdeva et al. conducted an industrial case study over nine months and proposed a novel approach to handling performance NFRs and security for big data and cloud-related projects using Scrum. The results show that their approach helps deal with performance and security requirements both individually and accounting for conflicts between them in an agile methodology~\cite{sachdeva2017handling}. Although relevant, this body of work has mainly focused on NFRs for traditional software systems (without the use of ML).

\textbf{RE for AI.} 
While there are many approaches  using ML to improve RE tasks such as model extraction~\cite{arora2019active}
or  prioritization~\cite{perini2012machine}, 
with much of such work reported in the AIRE Workshop Series~\cite{dalpiaz2020requirements}, 
 there is not as much work looking in the other direction, on RE for ML systems~\cite{vogelsang2019requirements}.
 However, recent papers point out challenges and issues looking at RE for AI-based systems.   Vogelsang \& Borg have pointed out that the ML-based development process becomes more complex, with the need to effectively use large quantities of data, as well as a dependence on other quality requirements (NFRs)~\cite{vogelsang2019requirements}.  Belani et al. identified, discussed, and tackled the challenges for requirement engineering disciplines in developing ML and AI-based complex systems~\cite{belani2019requirements}. They reported that one of the challenges in ML-enabled software development is to identify NFRs throughout the software  lifecycle, not only in the initial phases dealing with requirements, but as part of the whole lifecycle.   
Heyn et al. used three real use cases of distributed deep learning to describe system engineering challenges relating to requirements engineering~\cite{Heyn2021}.  They specifically focus on challenges concerning AI context, defining data quality attributes, testing/monitoring/reporting, and human factors.

Other papers begin to offer solutions, for example, Rahimi et al. introduced a RE-focused method using domain-specific concepts to find dataset gaps for safety-critical ML components~\cite{rahimi2019toward}.  Further work looking at requirements for AI focuses on specific types of requirements, such as transparency (e.g.,~\cite{felzmann2019transparency}) or legal requirements (e.g.,~\cite{bibal2020legal}). A recent workshop (RE4AI@REFSQ) has begun to explore RE for AI, but thus far, no papers have focused on the broad state of NFRs in AI industry.
Nakamichi et al. focused on quality characteristics and measurement methods related to functional correctness and maturity of ML software systems (MLS). They extended the quality characteristics of conventional software defined by ISO 25010 to those unique to MLS, defining a quality measurement method. They defined a method to identify requirements to derive the quality characteristics and measurement methods for MLS~\cite{nakamichi2020requirements}. 
Further work has focused on outlining the challenges of NFRs for ML-based solutions, including an outline of research directions. In contrast to work in this section, we focus on a wider perception of NFRs for ML in industry~\cite{horkoff2019non}. 

\textbf{SE for AI.}
Work has looked more broadly at how SE knowledge can be applied to AI and ML system development. Previous work in collaboration with companies has identified software engineering challenges of deep learning~\cite{arpteg2018software}. The research used seven ML projects and identified twelve challenges categorized into three areas: development challenges, production challenges, and organizational challenges. 
An empirical investigation on a taxonomy of SE challenges for ML systems has been presented and discussed in~\cite{lwakatare2019taxonomy}. 
Challenges include problem formulation and specifying the desired outcome, use of a non-representative dataset, lack of well-established ground truth, no deployment mechanism, and difficulties in building a highly scalable ML pipeline. 

Siebert et al. presented a quality model (i.e., quality aspects and evaluation objects) for software-intensive systems containing one or more components that use ML in an industrial use case~\cite{siebert2020towards}. 
Martínez-Fernández et al. conducted a systematic mapping study to collect and analyze current state-of-the-art knowledge about Software Engineering for AI-based systems.  They found that the most studied properties of AI-based systems are safety and dependability~\cite{martinez2021software}.

Further work on SE for AI can be found in workshops such a WAIN (Workshop on AI Engineering); 
however, most  papers found in general SE venues do not focus specifically on existing requirements challenges, and we are unable to locate papers which provide a broad, industrial view of NFRs.

\textbf{NFR Scope.} Work by Sculley et al.  focused on hidden technical debt in ML, but also pointed out that ML makes up only a small part of the system~\cite{sculley2015hidden}.  The ML component may be surrounded by code focusing on configuration, data collection, feature extraction, analysis tools, or monitoring, as well as glue code to make everything work together.  Further, as emphasized by Vogelsang \& Borg, RE for ML should focus not only on requirements for the system, but on requirements over the data~\cite{vogelsang2019requirements}.  These consideration raises the question of scope in our investigation.  To simplify, in this work we focus on three possible scopes:  NFRs over the ML model; NFRs over the whole system, including all the additional surrounding software as described by Sculley et al.; and NFRs for the ML-related data, as highlighted by Vogelsang \& Borg.

Overall, although  previous work has pointed out that handling NFRs in the development of ML systems is difficult,  little work focuses specifically on NFRs, or  tries to understand the state-of-the-art in  handling and treating  NFRs in industry.

\vspace{-0.1cm}
\section{Research Questions} \label{Research Questions}


To guide our study, we introduce a number of research questions. Our overarching research question is as follows:

\begin{itemize}[leftmargin=3em]
    \item[RQ1.] What is the perception and current treatment of NFRs in ML in industry?
\end{itemize}

We refine this RQ1 into more detail as follows:

\begin{itemize}[leftmargin=3em]

  \item[RQ1.1.]  Which ML-related NFRs are more or less important in industry?
 \item[RQ1.2.]  Over what aspects of the system are NFRs defined?
  \item[RQ1.3.]  What NFR and ML-related challenges are perceived?
  \item[RQ1.4.]  How are NFRs related to ML currently measured?
  \item[RQ1.5.]  Over what aspects of the system are NFRs measured?
  \item[RQ1.6.]  How are NFRs and their measurements captured in practice?
  \item[RQ1.7.]  What measurement-related challenges for NFRs in an ML-context exist?
  \end{itemize}

With RQ1.1, we aim to understand if the emphasis on certain NFRs in the literature corresponds with interest in reality.  RQ1.2 is inspired by our scoping question, are NFRs defined over the ML model, the whole system, or the data?  RQ1.3 aims to collect general challenges in this area.  In the final four RQs, we aim to understand if and how NFRs for ML are measured, over what part of the system they are measured, how these measurements are captured, and what challenges exist in the area of NFR for ML measurement.  
\vspace{-0.2cm}
\section{Methodology} \label{Methodology}


We choose a qualitative interview study to answer our research questions and explore experiences and perceptions in the context of NFRs for ML, following the ACM SIGSOFT Empirical Standard  (\url{tinyurl.com/QualitativeSurveys}).  

 \textbf{Sampling.}  The goal of the sampling was to find interviewees who had experience with ML, and who were currently working with ML in industry. 
 The sampling method was a mix of convenience, purposive, and snowball sampling.  We sent open calls to our colleagues and at industry events to find those with industrial ML experiences, then asked interviewees if they knew of further qualified people we could contact.  In the end, we interviewed 10 engineers and researchers in different sectors who have experience working with  ML in  industry. We believe the interviewees we selected are representative of those working in the data science and ML field, including their knowledge (or lack thereof) of NFRs.
 The interviews lasted 30 to 35 minutes, and were conducted online between December and February 2021 via Zoom. We recorded all the interview sessions with the permission of interviewees.  All interviews were  transcribed, and anonymized for further analysis.

\textbf{Data collection.}
We used  semi-structured interviews, with a set of predetermined open-ended questions, so that there remained enough freedom to add follow-up questions to collect in-depth information. 
The interview guide can be found in \ref{Table:interviewquestions}.  The interview started with describing the background of the study and the research gap to help interviewees to gain a sense of their role and make them comfortable with the context of the interview. The questions were divided into three categories. In the first set of questions (Questions 1-5), we collected interviewees' demographic information as well as their experience working with NFRs. In the second set of questions, questions 6 and 7 gather the interviewees' general impressions of NFRs, and if there are differences between NFRs for ML or traditional software.  These questions were meant to act as an initial check or filter:  if the participants did not believe NFRs were important, or that ML brought specific differences, the rest of the interview may not provide fruitful results. We then asked about NFR-related questions in an ML context, NFR measurements in an ML context, with a final open question for more input. For identifying and defining NFRs for ML-enabled software, we relied on the interviewees' own definition of NFRs, and reporting on their perceptions.

In some cases, participants were not explicitly familiar with the term NFR.  We believe this is not uncommon for those working with ML in industry.  In these cases we showed them an example hierarchy of NFRs, using McCall's quality hierarchy as an example~\cite{cavano1978framework}.  They were then able to recognize and talk about software quality aspects.  We discuss this as part of our consideration of validity threats in Sec.~\ref{sec:validity}.  

\begin{table}[]
    \centering
    \caption{Interview Questions Mapping to Research Questions}
    \vspace{-0.2cm}
    \label{Table:interviewquestions}
\begin{tabular} {|m{ 6.7cm}|m {1.2cm}|}
\hline
\textbf{Interview Questions}     &  \textbf{Research Questions} \\ \hline
\multicolumn{2}{|c|}{\textbf{Background of Interviewee (Demographic Data)}} \\ \hline  

1.	Please introduce yourself and your role in this company/ organization. &  N/A  \\ \hline
2.	Do you consider yourself more of academic person or industry related person? &  N/A  \\ \hline
3.	Total years of experience in the industry and how long you are in current position? & N/A   \\ \hline
4.	Please describe your responsibilities (e.g., Product owner, developer, Software Architect) in your organization. &  N/A  \\ \hline
5.	Please describe your experience working with NFRs. &  N/A  \\ \hline

\multicolumn{2}{|c|}{\textbf{NFR Related Questions}}    \\ \hline 

6.	Do you think NFRs play an important role for the success of a software? If yes, how? &  RQ1.1  \\ \hline
7.	Do you think there are differences in NFRs between generic software (without ML) and ML-enabled software?  If so, what?   If not, why not?  & RQ1.1    \\ \hline
8.	Do you think there are NFRs which are more prominent or important in ML context? If so, which ones? & RQ1.1   \\ \hline
9.	Do you think there are some NFRs less important in ML context which were prominent in generic software engineering?  &  RQ1.1  \\ \hline
10.	Do you think of NFRs for the whole system, for the ML model, for the data or other parts? &  RQ1.2  \\ \hline
11.	What challenges do you experience with NFRs for ML?  &  RQ1.3  \\ \hline

\multicolumn{2}{|c|}{\textbf{NFR Measurement Questions}}  \\ \hline 

12.	Do you measure NFRs over ML-enabled software?  &  RQ1.4  \\ \hline
13.	Of the NFRs you mentioned, how do you measure these NFRs in an ML context?  &  RQ1.4  \\ \hline
14.	Are these NFRs measured over the whole system, the ML implementation or some part of data?  &  RQ1.2  \\ \hline
15.	How do you capture NFRs and their measurement for ML-enabled systems?  &   RQ1.5 \\ \hline
16.	What are the challenges you face measuring NFRs for ML?  &  RQ1.6 \\ \hline
17.	Do you have anything else you would like to add?  & any   \\ \hline

\end{tabular}
\vspace{-0.5cm}
\end{table}

\textbf{Pre-Testing.}  To improve the validity and reliability of the interview process, we conducted test interviews with two Ph.D. students working with NFRs and ML. This procedure helped to remove ambiguous and redundant questions, revise unclear wordings and rearrange the questions. 

\textbf{Data Analysis.} The collected data were qualitative; therefore, we used thematic analysis as a data analysis method~\cite{runeson2009guidelines}.  We used a mixed form of coding, where we started with a number of high-level codes based on our RQs, then refined and adapted these codes when going through the transcripts~\cite{creswell2017research}.
Two authors started to code each transcribed interview separately and afterward reviewed and validated the codes for each interview with each other. We did this for the first five interviews and discussed the results and findings in several iterative meetings, reaching a good level of agreement.  Then the first author coded the remainder of the transcribed interviews. 
We then combined data from all transcriptions into summary tables and figures, working together to find high-level categories for our codes. We made an effort to maintain the original terminology of the participants in developing our codes, e.g., we merged similar NFRs only a few clear cases.  

As an example, the statement by P10,  
\begin{displayquote}``In terms of explainability, fairness, and other metrics, quality attributes, of course, it's a very important part of making any software as a service better. 
\end{displayquote}
is coded as \textsf{ImportantNFR}, as the interviewee discusses a number of important NFRs, and with more specific codes of \textsf{NFRCorrectnessAccuracy}, \textsf{NFRExplainability}, and \textsf{NFRFairness}, capturing the specific NFRs which arose.  

Interview transcripts and codes can be found here https://tinyurl.com/6d96vesb. 

\vspace{0.1cm}
\section{Results} \label{Results}

In this section, we provide our findings from the interviews.  We summarize our qualitative codes to provide an overview of NFRs in an ML context, then summarize interview demographics,  examine the reported importance of NFRs, the perceived challenges related to NFRs in ML, and measurement-related findings for NFRs in ML.
\vspace{-0.1cm}
\subsection{Overview: Thematic Codes}

\begin{figure}
    \centering
      \includegraphics[width=.5\textwidth]{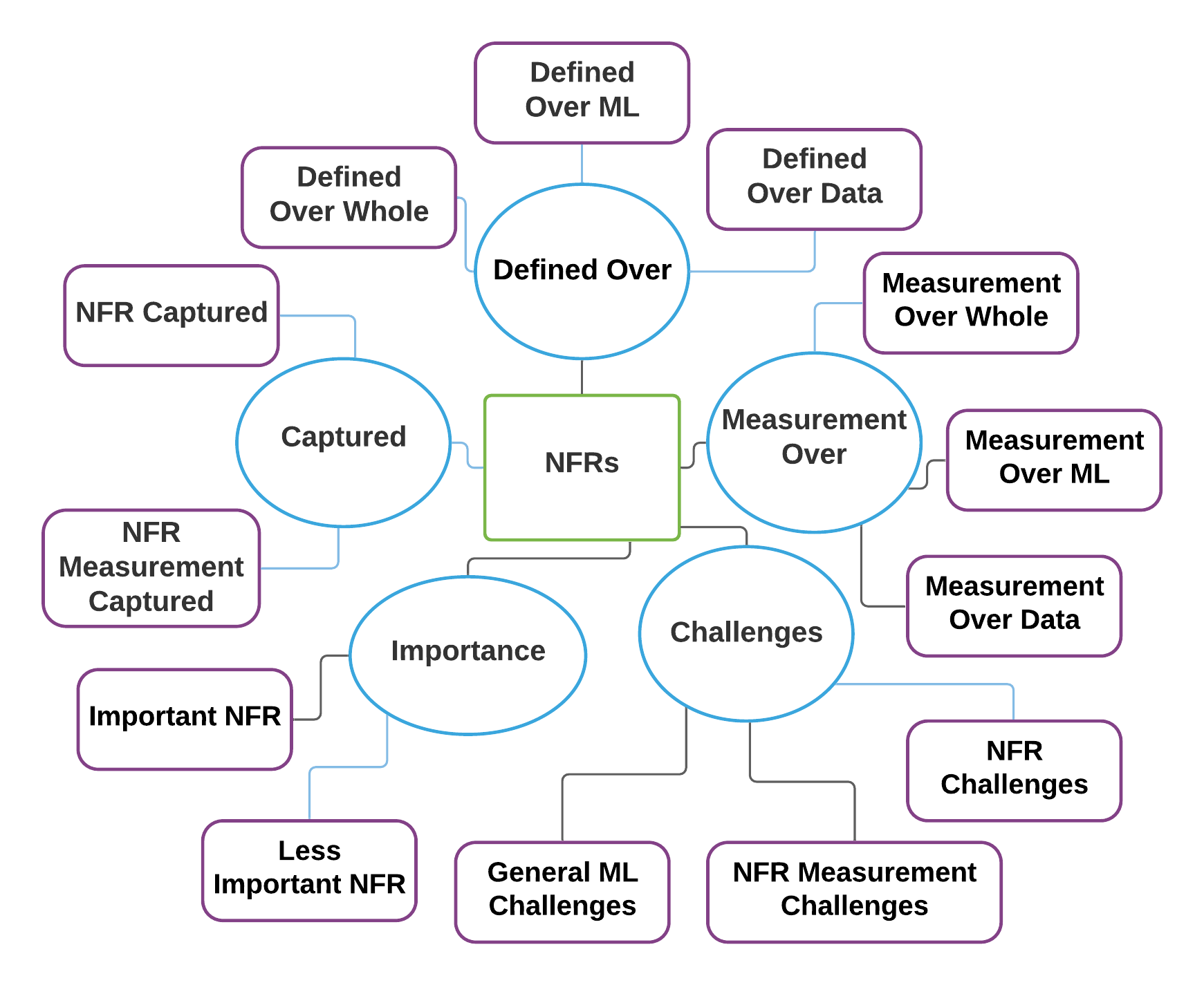}
    \vspace{-0.6cm}
    \caption{Overview of Codes/Themes}
    \label{fig:themes}
    \vspace{-0.5cm}
\end{figure}

In this section, we provide an overview of our thematic codes as derived from the interviews. 
A graphical overview can be found in Fig.~\ref{fig:themes}.
Five high-level codes were identified, sub-divided into different codes, e.g., \textsf{Challenges} is a high-level code divided into three sub-codes: \textsf{NFRChallenges}, \textsf{GeneralMLChallenges} and \textsf{NFRMeasurementChallenges}. 

The high-level code \textsf{DefinedOver} maps the interview comments that state which part of the system NFRs is defined over, subcategorized into: \textsf{DefinedOverData}- includes statements where interviewees said NFRs are defined over data, \textsf{DefinedOverML}- the interviewees said NFRs are defined over ML model, \textsf{DefinedOverWhole}- the interviewees said NFRs are defined over the whole software. For example, P4 said: 
\begin{displayquote}
“To be honest, I just see the non-functional requirements just for the machine learning system.”
\end{displayquote}
We mapped this comment with \textsf{DefinedOver} and \textsf{DefinedOverML} as this statement describes in which part of the system the NFRs should be defined. 

Similar to \textsf{DefineOver}, \textsf{MeasurementOver} mapped the statements that include comments on which part of the system NFRs are measured and categorized into \textsf{MeasurementOverData}, \textsf{MeasurementOverML}, \textsf{MeasurementOverWhole}. 

The statements on more and less important NFRs for ML are coded as \textsf{ImportantNFR} and \textsf{LessImportantNFR}, respectively.  Statements which include the name of specific NFRs for ML are coded with  their name, 
for example, Safety, performance, and efficiency 
are coded as \textsf{NFRsafety}, \textsf{NFRPerformance}, and \textsf{NFREfficiency}. 
 The methods for NFR and measurement capture were coded as \textsf{NFRCaptured} and \textsf{NFRMeasurementCapture}, respectively.

\vspace{-0.1cm}
\subsection{Participant Demographics}

We collected our interviewees' demographic information, asking them to introduce themselves, their role in their organization, their total years of experience and experience in their current position, their responsibilities, and their experience working with NFRs. The result is shown in Table~\ref{Table:demographics}.

\begin{table*}[]
\centering
\caption{Interviewees' Demographic Information}
\vspace{-0.2cm}
\label{Table:demographics}
\begin{tabular} {|p{0.8cm}|p{1cm}|p{1.5cm}|p{1.8cm}|p{0.8cm}|p{1cm}|p{3.6cm}|p{3.6cm}|}

\hline
\textbf{Partici-pant} & \textbf{Country} & \textbf{Domain}  & \textbf{Role} & \textbf{Exp. (Years)} & \textbf{Exp. in Position (Years)} & \textbf{Responsibilities} & \textbf{NFR Experience} \\ \hline
P1 & Sweden
&  Medical        &   Section Leader   &   20 & 1.5    &     Research and innovative projects on strategic level &  Consider NFRs while developing patient management tool     \\ \hline
P2 & Norway &  email service     &  Requirement Engineer and Tool Expert   & 11    &  1.5  &      Working with requirement engineering process, setting up tools to support process    &   Working with requirement engineering process but not with requirements themselves    \\ \hline
P3 & Canada &   Cloud Migration      &  Development Manager    &  15  &    2              &   Leading the team of software developers               &  Consider NFRs when creating and designing software              \\ \hline
P4 & Sweden
&  ML-Consulting       &  Senior Machine and Data Scientist     &   12 &  2    & Develop and implement algorithms for ML, co-ordinate development team  &  Work with NFRs              \\ \hline
P5 & Sweden
&    e-healthcare    & Chief Data Scientist     & 10    &     3           &         Lead team who are working for digitization and decision support system related to hospital         &  We use NFRs but not in very structured way              \\ \hline
P6 & Sweden
&  Sustainable solutions      &  Competence Team Manager    &   15  &    2           &     Designing and writing software, leading the project   &    Consider NFRs while designing and developing the software            \\ \hline
P7 & Sweden
&  Biotechnology  &      Consultant    &   3   &  1.5         &     Product owner, architect, developer                &      Consider NFRs while designing and developing the software          \\ \hline
P8 & Israel &  Cybersecurity      &     Head of Research  &  28    &  1              &   Take care of the research and supervise the development of the model               &   Consider NFRs and discuss during development of a product             \\ \hline
P9 & Sweden
&  Automotive  &     Functional safety expert           &   3   &  0.5        &   Researcher, function development, providing functional safety knowledge               &    Ensure safety related NFRs            \\ \hline
P10 & Canada & Multinational and multipurpose   &     Research and Technology lead        &   4   &  2            &    Lead a group of software engineers to make ML explainabile and accountable              &     Try to improve ML models considering NFRs           \\ \hline
\end{tabular}
\vspace{-0.4cm}
\end{table*}

Our interviewees cover a wide range of domains, industries and  countries working with different roles, e.g., Section leader, Data Scientist, Team manager, Consultant.  6/10 of the interviewees are from Sweden. The interviewees' responsibilities include conducting research, developing and implementing ML algorithms, leading the development team, etc. The interviewees' experience varies between 3 to 28 years, but most interviewees' had more than 12 years general experience. Overall, our interviewees had a leaning to more senior positions. However, on average they have only a few years experience in their current role.

We observed 
that the interviewees have a mix of industrial and research backgrounds,  it was hard to find people to speak to who are from the only industry without academic involvement. 
This could be a result of our search strategy, but believe this is may be an indication of the novelty of ML applications in industry. 
\vspace{-0.1cm}
\subsection{NFR Results}
To start, all the interviewees indicate that NFRs play important roles in the overall success of the software and that there were differences worth considering for ML. Thus we were able to continue with the rest of our questions.

\begin{figure*}
    \centering
    \includegraphics[width=1.0\linewidth]{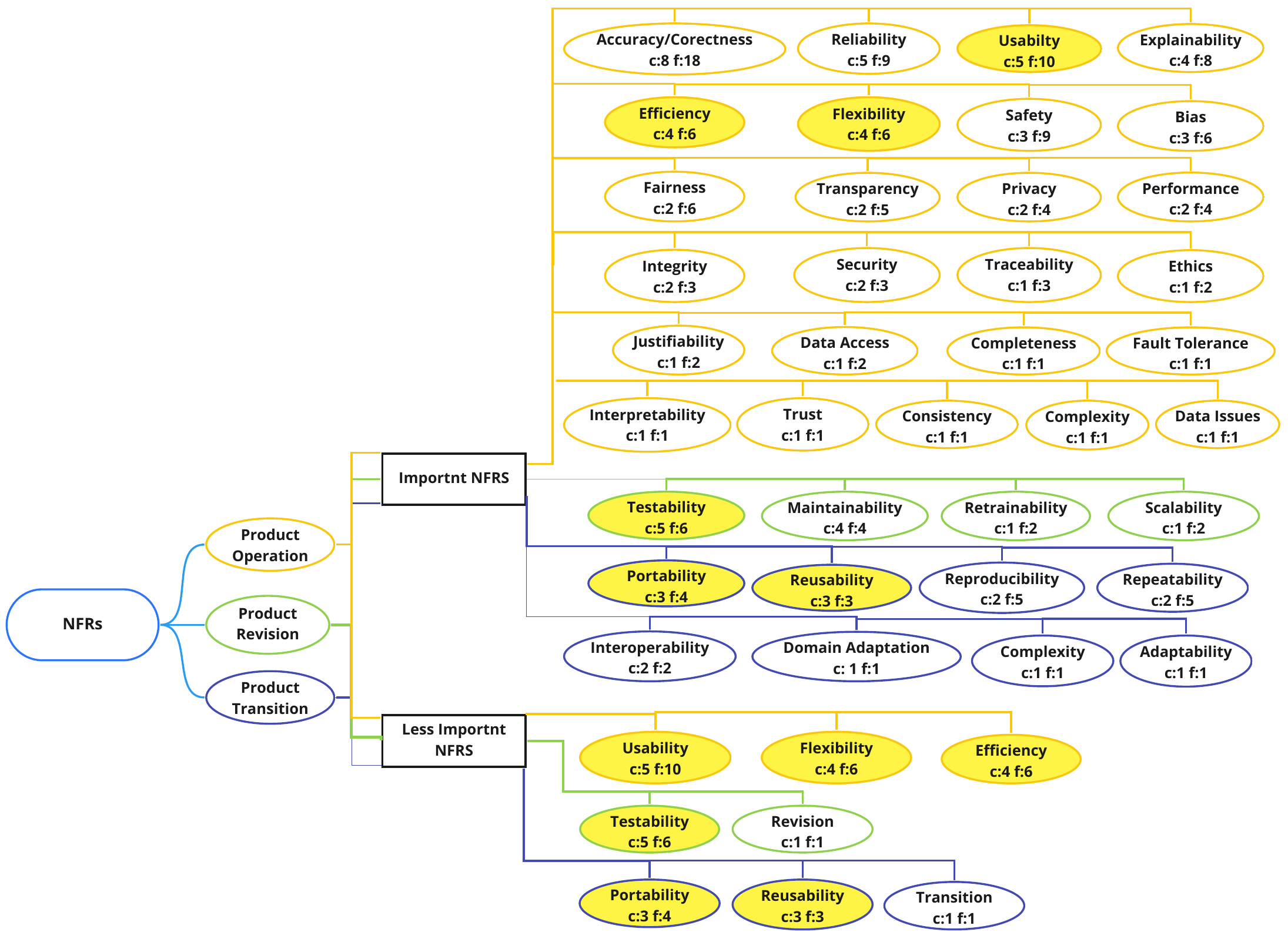}
    \caption{Important and Less Important NFRs for ML. \textbf{c:} counts of the number of the interviewees whose interview included, \textbf{f:} count of occurrences of the code across all transcripts, \textbf{Yellow background:} NFRs mentioned by some interviewees as important are identified as less important NFRs by other interviewees.}
    \label{fig:importance}
    \vspace{-0.4cm}
\end{figure*}

\noindent\textbf{Perceived NFR Importance (RQ1.1).} According to the interview codes, we identified important and less important NFRs for ML, categorizing these into the categories provided by~\cite{cavano1978framework}: product operation, revision, and  transition. The codes related to important and less important NFRs, counts of the number of the interviewees whose interview included the code (c), and the frequency (f), a count of occurrences of the code across all transcripts, are shown in Fig.~\ref{fig:importance}. We include the numbers to give an idea of frequency and ranking, however, given the small sample size, this ranking may change with more participants.  We are currently working on presenting and revising the ranking with more participants via a survey.  

All ten interviewees brought up important NFRs for ML-enabled software, and nine interviewees named less important NFRs. Interviewees could identify important NFRs for ML quickly compared to less important NFRs. 

While talking about important NFRs for ML, P3 named a number of NFRs:  
\begin{displayquote}
``Repeatability, accuracy, these things are often important in ML or deep learning-based software which is not generally that much present in traditional software.''
\end{displayquote} 
Concerning new NFRs, P4 stated:  
\begin{displayquote}``Retrainability is a new non-functional requirement for the Machine Learning system. When to retrain, how to retrain, which data use to retrain those are the requirements those you don't define in traditional software.''
\end{displayquote} 

\noindent Considering less important NFRs for ML, P1 commented,
\begin{displayquote}
``Flexibility right now is not so important. If you need to scale up, you can do some changing, so we don't consider that as much important thing yet. The same with reusability. I think as AI is not so much mature yet, so we are not considering it yet".
\end{displayquote}

The results illustrate that most NFRs are still considered important in an ML context, and few NFRs are considered less applicable. It is also important to note that there was a disagreement among the interviewees on which NFRs are less important. A few NFRs mentioned by some interviewees as less important are identified as important NFRs by other interviewees (colored yellow in Fig.~\ref{fig:importance}). Most of the interviewees could provide answers to the related interview questions. However, not everyone could answer this question easily, and they had to be shown a standard NFR hierarchy (McCall's) ~\cite{cavano1978framework} to illustrate possible NFRs.  \\
\\
\noindent\fbox{%
    \parbox{\columnwidth}{%
        \textbf{RQ1.1:  Most NFRs as defined for traditional software are still relevant in an ML context, while only a few become less prominent. Prominent NFRs according to our interviews include fairness, but also flexibility, usability, accuracy, efficiency, correctness, reliability, and testability.}
    }%
}\\

      

    
    



\noindent\textbf{Scope of NFRs (RQ1.2).}
We summarize our answers and codes regarding what part of the system NFRs are defined over:  the ML model, the data, or the whole system.  

Out of ten interviewees, eight said NFRs are defined over the ML model. As an example quote, P4 said:
\begin{displayquote}
“To be honest, I just see the non-functional requirements just for the machine learning system”.
\end{displayquote}
Two interviewees said NFRs are defined over the data, while 
four participants said NFRs are defined over the whole software. As an example, P7 mentioned:
\begin{displayquote}
“Machine learning projects also software project. So, I guess they all match all over.'' 
\end{displayquote}

The result shows that the NFRs for ML are mostly defined over the ML model or the system as a whole.  However, we see some disagreement here, and we  note that this question was not so easy to answer for many participants.\\ \\
\noindent\fbox{%
    \parbox{\columnwidth}{%
        \textbf{RQ1.2 (NFRs): Most participants focused on defining NFRs over the ML model itself or the whole system, and have not explicitly considered NFRs for ML-required data.}
    }%
}\\

\begin{figure*}
    \centering
    \includegraphics[width=1.0\textwidth]{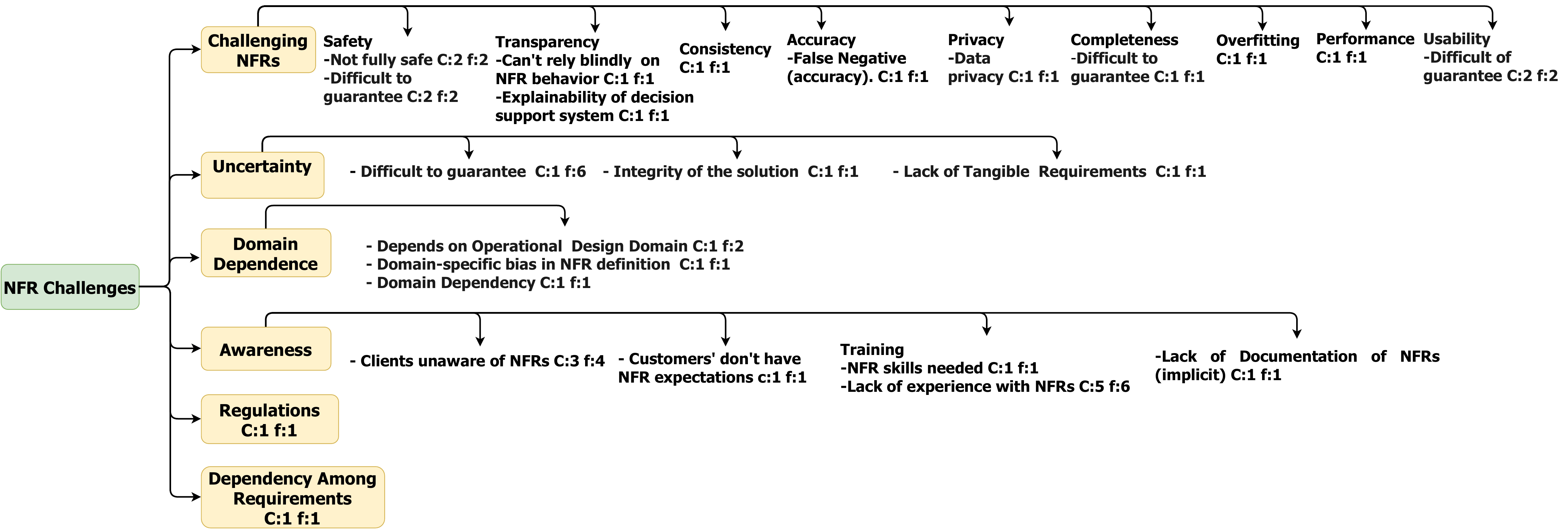}
    \vspace{-0.5cm}
    \caption{NFR-Related Challenges with ML Systems.}
    \label{fig:nfrchallenges}
    \vspace{-0.4cm}
\end{figure*}

\noindent\textbf{Reported NFR and ML-related Challenges (RQ1.3).}
All ten interviewees identified NFR-related challenges. These challenges are presented in Fig.~\ref{fig:nfrchallenges}.
Leaf-level challenges include interviewee counts (c) and frequencies (f).

According to the interviewees, NFR challenges include uncertainty, domain dependence, awareness, and regulations, illustrated in Fig.~\ref{fig:nfrchallenges}. The interviewees also brought up specific NFRs as challenges.  For example, some ML applications can be safety-critical, not entirely safe, and we can't rely blindly on NFR behavior (Safety). According to P2: 
\begin{displayquote}
``I think that's very tricky; you can really mess with safety. I think that's why companies always afraid of using machine learning techniques over traditional system where you can really check that.”
\end{displayquote}

Interviewees pointed out that the transparency of a decision support system can be crucial for sensitive ML applications, and while measuring accuracy, false negatives can be dangerous. According to the interviewees, maintaining the privacy and consistency of the system can also be very challenging.

We found several challenges related to uncertainty.  For ML-enabled software, it is challenging to guarantee pre-defined execution, preserve the integrity of the solutions, and there is a lack of tangible requirements. 
Other challenges related to domain dependence of NFRs.  Some NFRs for ML depend on, for example, an Operational Design Domain (ODD), which is the specific operating domain(s) in which an automated function or system is designed to properly operate, and there can be domain-specific bias in NFR definition. 

The responses also pointed out a lack of awareness of NFRs. Clients are often unaware of NFRs; therefore, they do not have expectations over NFRs. To define NFRs, special skills are needed, and the engineers and researchers  lack skills in this regard. P8 said:
\begin{displayquote}
``Then I think that we don't have enough experience in the field (NFRs for ML) to define them well".
\end{displayquote} 
The interviewees also mentioned a lack of proper documentation of NFRs, which made it more challenging to define NFRs for ML. 
Finally, at least one interviewee reported that regulations and laws constrain definition NFRs in ML-enabled software, and that this can be challenging.


When asked about NFR-related challenges in ML, some interviewees answered with both NFR-related challenges and more general challenges with ML. Eight interviewees described challenges not specifically related to NFRs.  For example, wrong training and testing data selection, complexity in data pre-processing, unexpected results over time, uncertain system behaviour, expensive and time consuming testing process, and unstructured development process.\\





\noindent\fbox{%
    \parbox{\columnwidth}{%
        \textbf{RQ1.3: Our interviewees reported many NFR-related challenges. Some NFRs were pointed out as particularly challenging (safety, transparency, accuracy, consistency, privacy, completeness), but other challenges included uncertainty, domain dependence of NFRs, a lack of awareness of NFRs and regulations.}
    }%
}









\subsection{NFR Measurement Results}

We collected NFR measurement-related codes in the third part of the interview, and present our findings in the following.

\textbf{NFR Measurements (RQ1.4).} While answering the question ``Do you measure NFRs over ML-enabled software?", all  interviewees answered they measure or need to measure NFRs over ML-enabled software. 
Answers to the question, ``Of the NFRs you mentioned, how do you measure these NFRs in an ML context?" varied depending on the functionalities the software provides. For example, NFRs can be measured based on response time, statistical analysis, different performance metrics, or user feedback. According to P10: 
\begin{displayquote}
``Lots of NFRs (e.g., accuracy, repeatability, consistency of execution, etc.)  are quantifiable, and those quantifiable NFRs can be measured by statistical analysis. For example, accuracy can be measured by accuracy matrix-like f1 score, root mean square error, etc.”
\end{displayquote}




 Measurement can be done by machine and human judgment combined, along with statistical analysis in safety critical sensitive measurements. P1 commented:
 \begin{displayquote}
 ``If you set up a clinical trial of something, then you compare with or without machine or with a doctor’s judgement with machine, then compare those and in the end if you do statistical analysis to see whether it is significant difference.” 
 \end{displayquote}
 
 Usability can be measured by interview results (P5):
 \begin{displayquote}
 ``We do perform interviews and use the result of the system and see how they find the usability.” 
 \end{displayquote}
 Additionally, usability can be measured by ad hoc methods, with some difficulty. P4 said: 
  \begin{displayquote}
  ``The usability of machine learning system is a bit tricky to measure, and sometimes you have to come up with this ad hoc matrix to know about  how usable the system is.”
\end{displayquote}

Further NFRs were also identified as challenging to measure, according to P10:
\begin{displayquote}
``Measurements should be done according to standard baseline, but some measurements are not quantifiable (e.g., usability, adaptability, flexibility, etc.), therefore tricky.''\\
\end{displayquote}
\noindent\fbox{%
    \parbox{\columnwidth}{%
        \textbf{RQ1.4: While some NFRs (e.g., accuracy) can be measured using ML-specific or more standard measures (e.g., precision, recall, f1 score), many are difficult to measure (e.g, fairness, explainability) as with traditional software.}
    }%
}\\

\textbf{NFR Measurement Scope (RQ1.5).} We summarize our results concerning what part of the system NFRs are measured: the data, the ML model, or the whole system. 
Three interviewees said NFRs for ML are measured over data (P8):
\begin{displayquote}
``Measurement is for the data. If you have labeled data for all cases, then you can measure the performance". 
\end{displayquote}
Six interview participants said NFRs were measured over the ML model, while 
four interviewees indicated measurements over the whole software.
P1 explained:
\begin{displayquote}
``Before you bring the system into production, you need to measure NFRs for the whole system.”
\end{displayquote}

  
  Generally, we see even more disagreement on the scope of measurement than on the scope of NFR definition, with still a slight focus on measuring over the model rather than the data or whole system.  \\
\\
\noindent\fbox{%
    \parbox{\columnwidth}{%
        \textbf{RQ1.5: As with definitions, there are variances in the scope of ML-related NFR measurements, with a focus on the ML model slightly favored.}
    }%
}\\

\textbf{NFR Measurement Capture (RQ1.6).}  
We asked the interviewees how NFR measurements for ML-enabled software were captured, e.g., in a tool, or via some documentation.  Many interviewees had difficulties answering this question, and we discuss this further in Sec.~\ref{sec:validity}. Some answered in terms of process. One respondent captures NFRs via interviews, while another mentioned use of checklists.
P8 said:
\begin{displayquote}
``I saw plenty of systems, and we still don't have a good enough methodology for that. Like, these are some checklists that you should go and do".
\end{displayquote}

For technical means to capture measurements, engineers use different methods. For example, they implement some algorithms that run and measure the result against time. According to P6:
\begin{displayquote}
``I think for this model, we should develop specific code. But we did not do it. My idea is that we have to write specific software to measure".
\end{displayquote}

One participant mentioned traceability tooling as a way to measure the fulfillment of NFRs. P9 stated:
\begin{displayquote}
``Well, normally we have one requirement tracing tool. So, if we have certain non-functional safety requirements, we define tests to prove that we fulfill this non-functional requirement".
\end{displayquote} 

In general, NFR measurement and measurement capture depend on the context. According to P1: 
\begin{displayquote}
``The measurement depends on their functionalities, some are time-based, and some are based on output. Sometimes measurement is captured using different tools and compared with journals in the field of healthcare".\\
\end{displayquote}
\noindent\fbox{%
    \parbox{\columnwidth}{%
        \textbf{RQ1.6: Interviewees were able to name some methods and tools to capture NFR measurements (e.g., checklists, custom code, traceability), but answers varied, and participants often found this question difficult to answer.}
    }%
}\\







\begin{figure*}
    \centering
    \includegraphics[width=1.0\textwidth]{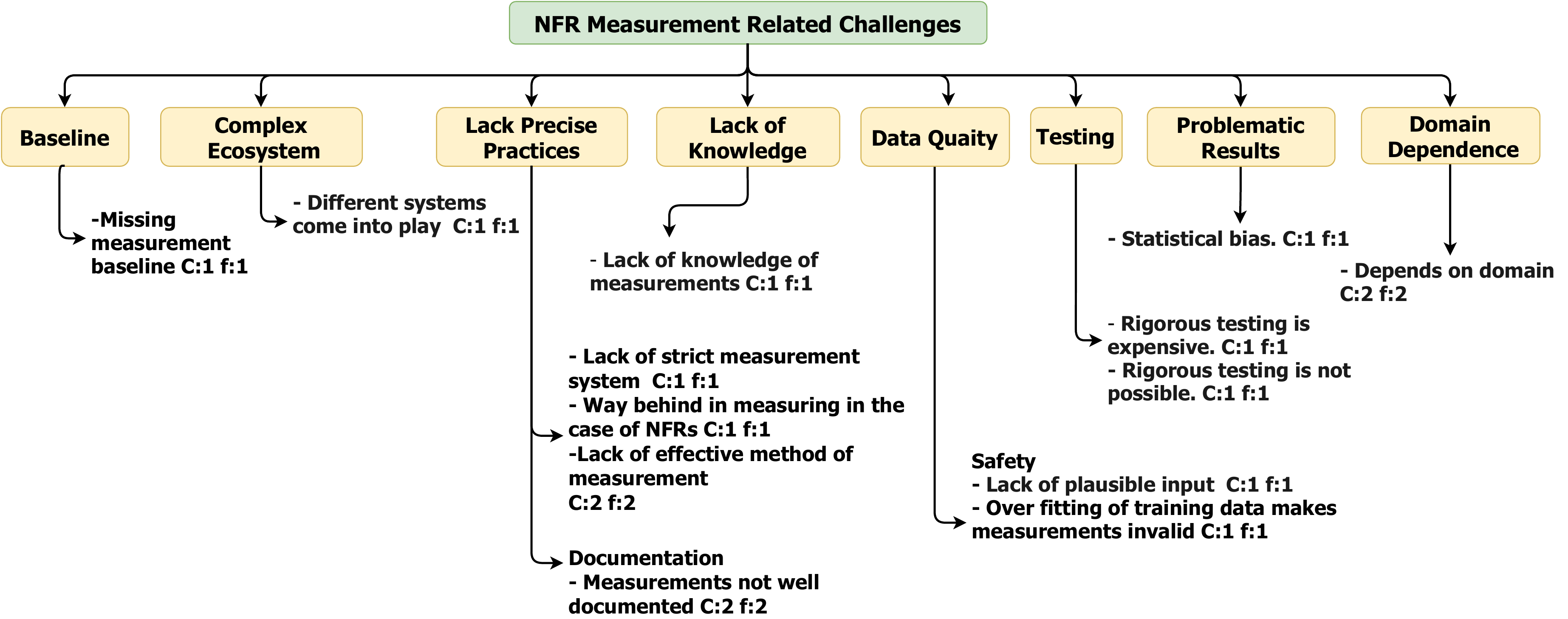}
    \vspace{-0.6cm}
    \caption{NFR Measurement-Related Challenges}
    \label{fig:measurementchallenges}
    \vspace{-0.5cm}
\end{figure*}

\textbf{NFR Measurement Challenges (RQ1.7).}  Fig.~\ref{fig:measurementchallenges} summarizes NFR measurement-related challenges as found via our coding process.
 Although many challenges could apply to both NFR challenges and NFR measurement challenges, e.g., domain dependence, the purpose is different. Here we mean the challenges specifically arise while measuring the NFRs. The first challenge concerns a missing measurement baseline and lack of a strict and effective measurement system. 
While measuring a certain NFR, different systems come into play. The engineers lack knowledge of measurements, and they are behind in measuring in the case of NFRs. According to P5: 
\begin{displayquote}
``If you compare the functional requirements, we are probably way behind when it comes to non-functional requirements. We do not have the same strict system for that as we do in the functional requirements." 
\end{displayquote}
In one case, a lack of knowledge of measurements is a challenge (P9):
\begin{displayquote}
``Not that I am aware of. I mean, testing the system of course, and based on those tests, we decide whether it is safe or not, but in Machine Learning, I am not aware of any possible measure yet".
\end{displayquote}

Further challenges include that the measurement of NFRs is domain-dependent and statistical bias exists. ML systems  depends on having plausible input, which can be difficult to find, and overfitting of training data makes measurements invalid. Participants also complained about the cost and plausibility of rigorous testing.


Lack of proper documentation on NFR measurements in the context of ML makes for further challenges. According to P4:
\begin{displayquote}
``Sometimes it is a lack of documents that contains non-functional requirements compatible with ML-enabled systems".\\
\end{displayquote}
\noindent\fbox{%
    \parbox{\columnwidth}{%
        \textbf{RQ1.7: Interviewees were able to list many challenges  measuring NFRs in ML-enabled systems, including a lack of knowledge, complexity, costly testing and finding data.}
    }%
}\\

\vspace{-0.2cm}
\section{Discussion} \label{Discussion}

In this section, we discuss our interview findings. 
Our results show that interviewees believe that NFRs for ML play a vital role in ML-enabled software's success. We were motivated to conduct our study to understand if the increased emphasis in the literature on certain NFRs such as fairness and transparency was reflected in industrial interest. We found that  our interviewees brought up many of these NFRs, including fairness, bias, and explainability. However, interviewees also discussed many other, more typical NFRs such as safety, interoperability, and retrainability. NFRs relating to performance, accuracy, and correctness were, unsurprisingly, emphasized by our interviewees. Furthermore, some NFRs without so much emphasis in the literature made an appearance, such as retrainability, justifiablity and testability. Few NFRs were agreed to be less important. To summarize, all the old NFRs we have dealt with for years are still very important for ML-enabled systems.  However, we have many further NFRs which have increased importance. 


The result also indicate that both engineers and customers lack knowledge and expertise in regards to NFRs for ML. There are a lack of documentation, methods, and benchmarks to define and measure NFRs for ML-enabled software.  Thus, this is a new area of exploration for our interview participants.

We have also found that there are varied answers for the scope of NFR definition in ML-enabled systems. Most participants are focusing on the ML model itself and not necessarily on the whole software, and there seems to be even less focus on NFRs over data, in contrast to recommendations found in~\cite{vogelsang2019requirements} and challenges identified in~\cite{Heyn2021}. The difference in results compared with ~\cite{vogelsang2019requirements} may be due to the difference in interviewees' profiles, with~\cite{vogelsang2019requirements} focusing on data scientists, while only 2/10 of our interviewees identified as data scientists. Our results can be seen to echo the findings in~\cite{belani2019requirements}:  although we did not ask specifically about NFRs in the software lifecycle, we found many measurement-related challenges related to system operation and testing.  

From a research perspective, our findings reveal several gaps that can shape future work. 

    1) We need further work which focuses on those NFRs with a newly increased focus in an ML context, e.g., fairness, explainability/transparency, bias, justifiability, and testability. This includes definitions, new taxonomies, measurements, and methods. Such work has already begun for some NFRs (e.g.,~\cite{brun2018software} for fairness, ~\cite{felzmann2019transparency} for transparency), but it is often approached from a general SE, rather than an RE perspective. 
    
    2) Conceptualization and methods are needed to address the scope of NFRs, e.g., there can different ways to view the sub-parts of the system, and these views may affect the way we categorize and define NFRs for ML~\cite{siebert2020towards}. We believe that many NFRs should be considered on three levels (ML model, software, data), but more work  must be done to investigate such ideas.
    
    3) Our findings have identified NFR-related challenges that can be addressed from an RE perspective by further research, e.g., the domain dependency of our NFR-ML understanding (as recently emphasized in~\cite{Heyn2021}), and awareness of NFRs for ML from the client and engineer side.

    4) New measurements for NFRs in an ML context are needed (e.g., work in~\cite{nakamichi2020requirements}) .  Many NFRs have always been difficult to measure, yet they must be measured in potentially new ways over different scopes.  The rise of ML makes measuring NFRs more difficult.
    
    5) We found further measurement-related challenges.  From an RE perspective, we can apply methods to understand complex ML ecosystems, to define and refine NFRs, or to make tradeoffs between NFRs (e.g., testing costs vs. results).  
    

From an industrial perspective, our findings provide a view of current practice, but do not yet offer concrete solutions.  However, it is useful for practitioners to see the sorts of questions and challenges that others are facing, and to understand that many of their current challenges are not unique.  

Overall, we see that this area is challenging for industry, yet important, and although individual companies may have some knowledge and practices, they do not have well established solutions for dealing with NFRs in an ML context.  


\textbf{Future Work.} 
This work has focused more on identification and importance of certain NFRs, we did not have time in the interviews to get into specific definitions, e.g., how is usability for ML-enabled software different than for traditional software. We are currently designing a survey to validate and refine our finding with more participants.  We are also working on a systematic literature review for NFRs in ML-based systems.  This will allow us to more precisely compare the focus of the literature to the focus in industry.  
Finally, we wish to investigate NFR scope as indicated above, identifying and differentiating NFRs over different system parts, such as data, the ML model, and the whole system.

\textbf{Threats to Validity. }\label{sec:validity} 
Our qualitative interview study has several threats to validity.  In terms of construct validity, we noted that questions concerning how NFR measurements were captured were not easy for the interviewees to understand. They could have understood each NFR differently.  
In retrospect, this question should  have been more clearly written, still we believe the results collected were interesting.

Several of our interviewees were not familiar with the concept of NFRs, and wanted examples. But we believe the interviewees are representative of the data science and ML field, and  many working in this area may not have software engineering training. Data science engineers are working to ensure the qualities of ML-enabled software, but they don't know traditional software engineering terms. To exemplify NFRs, we showed a version of McCall's software quality hierarchy as an example~\cite{cavano1978framework}, given its age and prominence, also used to categorize our NFR results.  We could have  used a number of other available NFR hierarchies, as there are many.  Reflecting on conclusion validity, showing a particular NFR hierarchy will bias the results towards that hierarchy; however, the differences between hierarchies is not extensive, and showing any one NFR hierarchy would influence the results in that direction. 

Our work suffers from the typical internal validity threats associated with thematic coding.  We mitigated these threats by performing independent coding over half the interviews and comparing results, finding sufficient agreement; and by using standard coding tools (NVivo) to help the process. We made the results available for further use.

In terms of the number of interviewees, we must consider whether we believe our findings were close to reaching saturation after 10 participants.  We found towards the end of our analysis that the codes were generally converging to a stable set.  However, the code ``justifiability'' was added in the last interview.  An eleventh interview, conducted, but not yet coded did not reveal any new results.  Thus, we believe further interviews could help to enrich our findings in a few areas, but would not produce significant additions.  

Our sampling technique found a number of participants who straddle the boundaries between industry and academia.  This may be a result of our circle of contacts, but we also believe that those who are interested in the topics covered in this paper are often mid- to upper-level management, and often have a strong academic background.  Our participants are mainly from one country, threatening external validity.  However, we found participants in a diverse set of industries, and we believe Sweden has a strong and international AI-oriented industry, thus our participants are fairly representative.

    
\vspace{-0.1cm}
\section{Conclusions}  \label{Conclusions}

We have conducted a qualitative interview study to understand the perception of and practices for NFRs in ML-based systems in industry. The interviewees agree that NFRs for ML is an important topic. The results showed that traditional NFRs like usability can  still be important for ML, as new NFRs such as transparency gain importance. The results also showed that NFRs for ML are not well-structured or well-documented. According to interviewees’ industrial and research experience, it is difficult to define and measure NFRs for ML as ML-enabled systems. People are not aware of NFRs for ML-enabled software, and engineers and researchers lack knowledge and experience. Sometimes the NFRs are not tangible, which creates measurement challenges and complex, expensive testing. From an industrial perspective, NFRs for ML are not well organized and well developed and their consideration is mainly in an initial stage. The challenges and complexities of NFR-related research remain, but are intensified by ML.  Much further research by the RE community is needed to overcome NFR-related challenges for ML-enabled software.

\section{Acknowledgements} \label{Acknowledgements}
\noindent This work is supported by a Swedish Research Council (VR) Project: Non-Functional Requirements for Machine Learning: Facilitating Continuous Quality Awareness (iNFoRM).

\bibliographystyle{IEEEtran}
\bibliography{references}

\end{document}